\documentclass{aa}  

\usepackage{graphicx}
\usepackage{mathptmx}
\usepackage[T1]{fontenc}
\usepackage{ae,aecompl}
\usepackage{amsmath}	
\usepackage{amssymb}	
\usepackage{natbib}
\usepackage{pdflscape}
\usepackage[para]{threeparttable}
\usepackage{placeins}
\usepackage{flafter}
\usepackage{multicol}
\usepackage{caption}
\usepackage{dblfloatfix}
\begin{document}

   \title{Formation of interstellar propanal and 1-propanol ice: a pathway involving solid-state CO hydrogenation}

   \author{D. Qasim\inst{1},
  G. Fedoseev\inst{2},
  K.-J. Chuang\inst{1,3}\thanks{Present address: Laboratory Astrophysics Group of the Max Planck Institute for Astronomy at the Friedrich Schiller University Jena,
Institute of Solid State Physics, Helmholtzweg 3, D-07743 Jena, Germany},
   V. Taquet\inst{4},
   T. Lamberts\inst{5},
   J. He\inst{1},
   S. Ioppolo\inst{6},\\
   E. F. van Dishoeck\inst{3},
          \and
          H. Linnartz\inst{1}
          }

   \institute{Sackler Laboratory for Astrophysics, Leiden Observatory, Leiden University, PO Box 9513, NL--2300 RA Leiden, The Netherlands\\
              \email{dqasim@strw.leidenuniv.nl}
              \and INAF--Osservatorio Astrofisico di Catania, via Santa Sofia 78, 95123 Catania, Italy \
              \and Leiden Observatory, Leiden University, PO Box 9513, NL--2300 RA Leiden, The Netherlands \
              \and INAF-Osservatorio Astrofisico di Arcetri, Largo E. Fermi 5, I-50125 Florence, Italy \
             \and Leiden Institute of Chemistry, Leiden University, PO Box 9502, NL--2300 RA Leiden, The Netherlands \
             \and School of Electronic Engineering and Computer Science, Queen Mary University of London, Mile End Road, London E1 4NS, UK
             }

   \date{Received X; accepted Y}

 
  \abstract
   {1-propanol (CH$_3$CH$_2$CH$_2$OH) is a three carbon-bearing representative of the primary linear alcohols that may have its origin in the cold dark cores in interstellar space. To test this, we investigated in the laboratory whether 1-propanol ice can be formed along pathways possibly relevant to the prestellar core phase.}
   {We aim to show in a two-step approach that 1-propanol can be formed through reaction steps that are expected to take place during the heavy CO freeze-out stage by adding C$_2$H$_2$ into the CO + H hydrogenation network via the formation of propanal (CH$_3$CH$_2$CHO) as an intermediate and its subsequent hydrogenation.}
   {Temperature programmed desorption-quadrupole mass spectrometry (TPD-QMS) was used to identify the newly formed propanal and 1-propanol. Reflection absorption infrared spectroscopy (RAIRS) was used as a complementary diagnostic tool. The mechanisms that can contribute to the formation of solid-state propanal and 1-propanol, as well as other organic compounds, during the heavy CO freeze-out stage are constrained by both laboratory experiments and theoretical calculations.}
   {Here it is shown that recombination of HCO radicals formed upon CO hydrogenation with radicals formed via C$_2$H$_2$ processing -- H$_2$CCH and H$_3$CCH$_2$ -- offers possible reaction pathways to solid-state propanal and 1-propanol formation. This extends the already important role of the CO hydrogenation chain to the formation of larger complex organic molecules (COMs). The results are compared with ALMA observations. The resulting 1-propanol:propanal ratio concludes an upper limit of $< 0.35 - 0.55$, which is complemented by computationally derived activation barriers in addition to the experimental results.}
   {}


   \authorrunning{Qasim et al.}
   \titlerunning{Formation of interstellar propanal and 1-propanol ice}
   \maketitle
%

\section{Introduction}
\label{introduction}

The search for three carbon-bearing aldehydes and alcohols has been the subject of a number of devoted observational studies. An example of recent observations of such species is the work by \citet{lykke2017alma}, where propanal (an aldehyde), among other organics, was detected towards the low-mass protostar IRAS 16293-2422B. In addition to these observations, propanal has also been identified in the Sagittarius B2 North (Sgr B2(N)) molecular cloud \citep{hollis2004green,mcguire2016discovery} and within the Central Molecular Zone of the Milky Way \citep{requena2008galactic}. Its detection on comet 67P/Churyumov-Gerasimenko was claimed by \citet{goesmann2015organic} but is still under debate \citep{altwegg2017organics}. Given the chemical link between aldehydes and alcohols, it is expected that propanol will be formed alongside propanal. Yet in comparison to propanal, the number of reported detections of 1-propanol in observational projects is very limited. Observations towards Sgr B2(N2), the northern hot molecular core within Sgr B2(N), only lead to an upper limit value of < $2.6 \times 10^{17}$ cm$^{-2}$ for 1-propanol \citep{muller2016exploring}. \citet{tercero2015searching} discussed the identification of 1-propanol towards Orion KL, but their claim has been questioned by others \citep{muller2016exploring}. The detection of propanol (without isomeric details) on comet 67P/Churyumov-Gerasimenko was reported by \citet{altwegg2017organics}.

In the laboratory, both propanal and propanol have been synthesized in astrophysical ice analogue experiments that require `energetic' processing for product formation. `Energetic' refers here to a radical-induced process that requires the involvement of UV, cosmic rays, and/or other `energetic' particles. \citet{kaiser2014infrared} and \citet{abplanalp2016study} showed that propanal can be formed by the electron-induced radiation of CO:CH$_4$ or CO:C$_2$H$_6$. \citet{hudson2017laboratory} were able to form propanal by proton irradiation of a CO$_2$:C$_3$H$_6$ ice mixture at 10 K. H$_2$O:$^{13}$CH$_3$OH:NH$_3$ 78 K ice exposed to UV photons and heated to room temperature also yielded propanal \citep{de2015aldehydes}. Propanol was reported to be formed by electron irradiation of a $^{13}$CO:$^{13}$CD$_4$ ice mixture at 5 K in experiments that did not allow to discriminate between 1- and 2-propanol \citep{abplanalp2018synthesis}.

In both the laboratory and observational work, propanal has been detected in conjunction with other organics such as acetone, propylene oxide, acetaldehyde, and so on. This demonstrates that propanal may be a reaction product in a number of astrochemical formation networks and its presence in the ISM may therefore be linked to the formation of a range of organic species. In this article, we focus solely on the formation of propanal and its direct derivative, 1-propanol, focusing on pathways relevant to the prestellar core, that is low temperature of $\sim$10 K and predominantly `non-energetic' processing. `Non-energetic' is used to refer to radical-induced processes that do not involve external energy input such as UV, cosmic rays, and/or electrons.   

The particular focus on 1-propanol is strongly motivated by the astrobiological relevance of this compound. 1-propanol is a primary alcohol, and it is hypothesized that primary alcohols may have been the constituents of cell membranes during abiogenesis. Cell membranes are currently and commonly composed of glycerophospholipids \citep{moran2012principles}, but whether such complex amphiphiles could be available on the early Earth is debated \citep{deamer2002first}. More simple and thus more likely lipids would be those composed of primary alcohols, such as prenol lipids. Additionally, the cell membranes of archaea (i.e., domain of ancient prokaryotic unicellular organisms) are known to be composed of primary alcohols \citep{de1986structure}, providing extra motivation to investigate formation routes of primary alcohols, including propanol.     

In this study we investigate whether propanal and 1-propanol can be formed by adding acetylene (C$_2$H$_2$) to the CO + H surface reaction chain. That is, we focus on the `non-energetic' (dense cloud relevant) processing of the ice. It has been experimentally demonstrated that complex organic molecules (COMs) -- as large as glycerol (a polyol compound) and/or glyceraldehyde (an aldose) -- can be formed below 20 K and without `energetic' input via the solid-state CO hydrogenation network \citep{fedoseev2015experimental,fedoseev2017formation,butscher2015formation,butscher2017radical,chuang2015h}. This aligns with the observationally constrained heavy CO freeze-out stage \citep{pontoppidan2006spatial,boogert2015observations,qasim2018formation}. It has been shown that the CO + H reaction product, formaldehyde (H$_2$CO), can be hydrogenated to form methanol (CH$_3$OH) \citep{watanabe2002efficient,fuchs2009hydrogenation}. In a somewhat related way, glycolaldehyde (HCOCH$_2$OH) and ethylene glycol (H$_2$COHCH$_2$OH) are proposed to be linked through sequential H-addition reactions \citep{fedoseev2017formation}. Additionally, acetaldehyde (CH$_3$CHO) can be hydrogenated to form ethanol (CH$_3$CH$_2$OH) \citep{bisschop2007h}. Thus we expect propanal to be hydrogenated to form 1-propanol. 

The hydrogenation of C$_2$H$_2$ has a barrier \citep{kobayashi2017hydrogenation} and it is expected that in space, hydrocarbon radicals formed by atom-addition are good candidates to combine with reactive CO + H intermediates to form COMs. For these reasons, in this study, the CO and C$_2$H$_2$ solid-state hydrogenation chains are connected to investigate the formation of reaction products that cannot be formed along the individual hydrogenation schemes. It should be noted that C$_2$H$_2$ has not yet been observed in interstellar ices. In the experiments discussed below, C$_2$H$_2$ was used both as a likely interstellar precursor species, and as a tool to form hydrocarbon radicals, in a comparable way to how O$_2$ was used to generate OH radicals \citep{cuppen2010water}. 
 
This paper is organised in the following way. Section~\ref{sect2} is an overview of the experimental setup and performed experiments. Section~\ref{results} contains results that show how propanal and possibly 1-propanol are formed by the simultaneous hydrogenation of CO and C$_2$H$_2$, and how propanal hydrogenation unambiguously results in the formation of 1-propanol. In Sect.~\ref{4}, we discuss the identification and formation pathways of a variety of organic compounds. Section~\ref{astro} is a discussion on how this combined laboratory work and theoretical calculations connect to the chemical inventory during the heavy CO freeze-out stage, and compares the outcomes with recent observations from the Atacama Large Millimeter/submillimeter Array (ALMA). Section~\ref{conclusions} is a summary of the findings presented in this paper.

\section{Experimental procedure}
\label{sect2}
\subsection{Description of the setup}
\label{sec:setupdescription}
\begin{table*}[btp!]
	\centering
	\caption{A list of the selected experiments and experimental conditions. Molecular fluxes were determined by the Hertz-Knudsen equation.}
	\label{table1}
	\begin{tabular}{c c c c c c c c c c} 
		\hline
		No. & Experiments & Ratio & T$_{\mathrm{sample}}$ & Flux$_{\mathrm{{C}_2{H}_2}}$ & Flux$_{\mathrm{CO}}$ & Flux$_{\mathrm{H}}$ & Flux$_{\mathrm{propanal}}$ & Flux$_{\mathrm{1-propanol}}$ & Time\\
 & & C$_{2}$H$_{2}$:CO:H & K & cm$^{-2}$s$^{-1}$ & cm$^{-2}$s$^{-1}$ & cm$^{-2}$s$^{-1}$ & cm$^{-2}$s$^{-1}$ & cm$^{-2}$s$^{-1}$ & s \\ 
		\hline
		1.0 & C$_{2}$H$_{2}$ + CO + H & 1:2:10 & 10 & $5 \times 10^{11}$ & $1 \times 10^{12}$ & $5 \times 10^{12}$ & - & - & 21600\\ 
        1.1 & C$_{2}$H$_{2}$ + CO & - & 10 & $5 \times 10^{11}$ & $1 \times 10^{12}$ & - & - & - & 21600\\ 
        1.2 & C$_{2}$H$_{2}$ + H & - & 10 & $5 \times 10^{11}$ & - & $5 \times 10^{12}$ & - & - & 21600\\ 
        1.3 & C$_{2}$H$_{2}$ + C$^{18}$O + H & 1:2:10 & 10 & $5 \times 10^{11}$ & $1 \times 10^{12}$ & $5 \times 10^{12}$ & - & - & 21600\\ 
        1.4 & C$_{2}$H$_{2}$ + $^{13}$C$^{18}$O + H & 1:2:10 & 10 & $5 \times 10^{11}$ & $1 \times 10^{12}$ & $5 \times 10^{12}$ & - & - & 21600\\ 
		2.0 & 1-propanol & - & 10 & - & - & - & - & $1 \times 10^{12}$ & 3600\\ 
		2.1 & propanal + H & - & 10 & - & - & $5 \times 10^{12}$ & $3 \times 10^{12}$ & - & 28800\\ 
        2.2 & propanal + H & - & 10 & - & - & $5 \times 10^{12}$ & $2 \times 10^{11}$ & - & 7200\\ 
        2.3 & propanal & - & 10 & - & - & - & $2 \times 10^{12}$ & - & 3600\\ 
		2.4 & propanal & - & 10 & - & - & - & $3 \times 10^{14}$ & - & 100\\ 
        
		\hline
	\end{tabular}
    \end{table*}

All experiments described in this study took place in the ultrahigh vacuum (UHV) setup, SURFRESIDE$^2$. The design of the setup is described by \citet{ioppolo2013surfreside2}, and details on the recent modifications are given by \citet{fedoseev2017formation}, \citet{chuang2018reactive}, and \citet{qasim2018formation}. Below, only the relevant settings are summarised. Ices were formed on a gold-plated copper substrate that is positioned in the centre of the main chamber (base pressure of low $\sim$$10^{-10}$ mbar range) and can be cooled to 7 K by a closed-cycle helium cryostat and heated to 450 K by resistive heating. Substrate temperatures were measured by a silicon diode sensor with a 0.5 K absolute accuracy.   

Connected to the central vacuum chamber are two atomic beam lines. Hydrogenation of the ice was possible by a Hydrogen Atom Beam Source (HABS) \citep{tschersich1998formation,tschersich2000intensity,tschersich2008design}. H-atoms were formed by the thermal cracking of hydrogen molecules (H$_{2}$; Linde 5.0) within the HABS chamber. As the atoms and undissociated H$_2$ molecules exited the HABS chamber, they were collisionally cooled by a nose-shaped quartz pipe before landing on the icy substrate, where they were thermalized instantly to the temperature of the substrate. The second atomic beam line, a microwave plasma atom source, was not used in the present study. 

Gases and vapours were prepared as follows. Acetylene (5\% of C$_2$H$_2$ in He; Linde 2.6) and carbon monoxide (CO; Linde 4.7) entered the main chamber via two separate pre-pumped dosing lines equipped with two leak valves. $^{13}$CO (Sigma-Aldrich 99\%) and $^{13}$C$^{18}$O (Sigma-Aldrich 99\%) isotopologues were used as tools to confirm the identification of the formed products. Propanal (Sigma-Aldrich $\geq$ 98\%) and 1-propanol (Honeywell $\geq$ 99.9\%) solutions, which were placed in individual glass tubes connected to the gas manifold by ultra-torr fittings, underwent freeze-pump-thaw cycles in order to remove gas impurities and were subsequently bled into the main chamber through the aforementioned dosing lines.  

Two complementary diagnostic tools were used to monitor ice processing. Reflection absorption infrared spectroscopy (RAIRS) simultaneously samples the consumption of precursor material and the formation of reaction products by visualizing the intensity decrease or increase, respectively, of molecule specific vibrational modes. In our setup, a Fourier Transform Infrared Spectrometer (FTIR) was used to cover the 4000-750 cm$^{-1}$ region with a spectral resolution of 1 cm$^{-1}$. In total, 512 scans were averaged over 230 seconds to obtain one spectrum. Temperature programmed desorption-quadrupole mass spectrometry (TPD-QMS) was used to investigate the thermally desorbed ice constituents as a function of desorption temperature. A typical ramp rate of 5 K/min was applied. The QMS electron impact source was operated at 70 eV, which induces well characterised and molecule specific fragment patterns. RAIRS is less sensitive than TPD-QMS, but has the advantage that it does not destroy the ice. The latter probes two molecule-specific parameters: the desorption temperature and the electron impact induced fragmentation pattern. In general, this combination allows unambiguous molecule identifications, particularly when isotopic species are also used as a cross-check. For an overview of the positives and negatives of both methods, see the work by \citet{ioppolo2014laboratory}. 

\subsection{Overview of experiments}
\label{experimentalmethods}

Table~\ref{table1} lists the experiments that were performed in this study. All fluxes were determined via the Hertz-Knudsen equation \citep{kolasinski2012} except for the H-atom flux, which was based on an absolute D-atom flux measured by \citet{ioppolo2013surfreside2}. The purpose of the experiments is described below.

Experiments 1.0-1.4 were used to verify the formation of propanal by the radical--radical recombination reaction between the radicals formed from hydrogenation of CO and C$_{2}$H$_{2}$. Experiment 1.0 was compared to experiments 1.1 and 1.2 to demonstrate that product formation requires radical species to be formed in the ice. We note that the listed C$_{2}$H$_{2}$:CO:H ratio in Table~\ref{table1} was experimentally found to be the most favourable ratio for product formation among our set of performed ratios (not discussed here). Carbon monoxide (CO) isotopologues were exploited in experiments 1.3 and 1.4 to witness the mass-to-charge (\emph{m/z\/}) shift in the TPD experiments that must occur if propanal (and 1-propanol) is formed.

Experiments 2.0-2.4 were used to verify the formation of 1-propanol ice via the surface hydrogenation of propanal at 10 K. Experiment 2.0 provides a 1-propanol reference. The TPD spectra of experiments 2.0, 2.2, and 2.3 were analysed to verify 1-propanol formation. Experiments 2.3 and 2.4 were used as controls to verify that the IR feature at 969 cm$^{-1}$ in experiment 2.1 does not overlap with the features of propanal. The feature was additionally compared to the IR spectrum of experiment 2.0.

It should be noted that in all experiments, the precursor species listed in Table~\ref{table1} were used in co-deposition experiments. These result in a higher product abundance compared to experiments in which pre-deposited precursor species are bombarded. Moreover, co-deposition is more representative for the actual processes taking place in space \citep{linnartz2015atom}.

\begin{figure*}[bpt!]
\includegraphics[width=\textwidth]{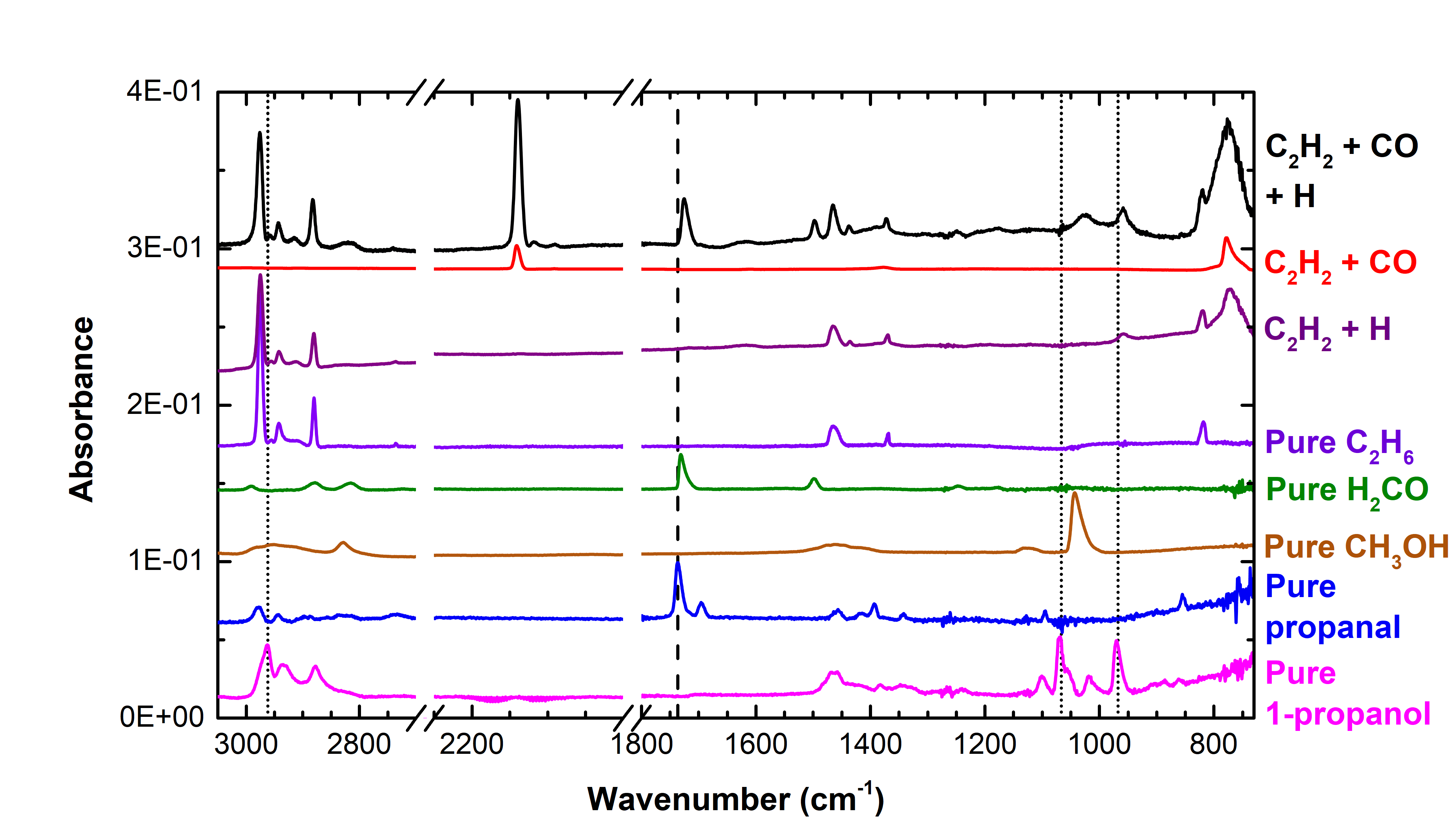}
\caption{RAIR spectra obtained after the deposition of C$_{2}$H$_{2}$ + CO + H (exp. 1.0), C$_{2}$H$_{2}$ + H (exp. 1.2), 1-propanol (exp. 2.0), CH$_3$OH ($5 \times 10^{15}$ cm$^{-2}$), propanal (exp. 2.3), H$_2$CO ($5 \times 10^{15}$ cm$^{-2}$), C$_2$H$_6$ ($5 \times 10^{15}$ cm$^{-2}$), and C$_{2}$H$_{2}$ + CO (exp. 1.1) on a 10 K surface. The spectrum of C$_2$H$_6$ is adapted from the work by \citet{oberg2009formation}. The dashed and dotted lines highlight the frequencies that correlate to the strongest features of propanal and 1-propanol, respectively. Spectra are scaled to highlight the IR features of interest, and are offset for clarity.}
\label{fig3}
\end{figure*}

\section{Results}
\label{results}

\begin{table*}[btp!]
\centering
	\caption{List of assigned IR absorption features in the co-deposition of C$_{2}$H$_{2}$ + CO + H (exp. 1.0).}
	\label{table2}
    \begin{threeparttable}
	\begin{tabular}{c c c c c} 
		\hline
        Peak position & Peak position & Molecule & Mode & Reference\\
(cm$^{-1}$) & ($\mu$m) \\ 
\hline
776 & 12.887 & C$_{2}$H$_{2}$ & $\upsilon_5$ & This work\\
820 & 12.195 & C$_{2}$H$_{6}$ and C$_{2}$H$_{4}$ & $\upsilon_{12}$ and $\upsilon_{10}$ & a,b,c,d,e,f,g,h,i\\
959 & 10.428 & C$_{2}$H$_{4}$ & $\upsilon_7$ & a,b,c,e,f,g,h,i,j \\
1025 & 9.756 & CH$_{3}$OH & $\upsilon_8$ & k,l \\
1371 & 7.294 & C$_{2}$H$_{6}$ & $\upsilon_6$ & a,b,d,e,f,g,h,i\\
1438 & 6.954 & C$_{2}$H$_{4}$ & $\upsilon_{12}$ & a,b,c,d,f,g,h,i\\
1466 & 6.821 & C$_{2}$H$_{6}$ & $\upsilon_{11}$ or $\upsilon_{8}$ & a,b,c,d,e,f,g,h,i\\
1498 & 6.676 & H$_{2}$CO & $\upsilon_3$ & k,l\\
1726 & 5.794 & H$_{2}$CO & $\upsilon_2$ & k,l\\
2138 & 4.677 & CO & $\upsilon_1$ & k,l\\
2882 & 3.470 & C$_{2}$H$_{6}$ & $\upsilon_5$ & a,b,c,e,f,g,h,i\\
2915 & 3.431 & C$_{2}$H$_{6}$ & $\upsilon_8$ + $\upsilon_{11}$ & e,c\\
2943 & 3.398 & C$_{2}$H$_{6}$ & $\upsilon_8$ + $\upsilon_{11}$ & a,b,c,d,e,f,g,h\\
2958 & 3.381 & C$_{2}$H$_{6}$ & $\upsilon_1$ & e,g\\
2976 & 3.360 & C$_{2}$H$_{6}$ and C$_{2}$H$_{4}$ & $\upsilon_{10}$ and $\upsilon_{11}$ & a,c,d,e,f,g,h,i\\
\hline
\end{tabular}
\begin{tablenotes}
\item[a]\citet{kim2010laboratory}
\item[b]\citet{zhou2014radiolysis}
\item[c]\citet{abplanalp2018untangling}
\item[d]\citet{gerakines1996ultraviolet}
\item[e]\citet{abplanalp2016complex}
\item[f]\citet{moore1998infrared}
\item[g]\citet{bennett2006laboratory}
\item[h]\citet{moore2003infrared}
\item[i]\citet{hudson2014infrared}
\item[j]\citet{kobayashi2017hydrogenation}
\item[k]\citet{watanabe2002efficient}
\item[l]\citet{chuang2015h}
\end{tablenotes}
\end{threeparttable}
\end{table*}

\subsection{Formation of propanal from C$_2$H$_2$:CO hydrogenation}
\label{3.2}

Figure~\ref{fig3} shows the RAIR spectrum obtained after the co-deposition of C$_{2}$H$_{2}$ + CO + H at 10 K. A list of the identified RAIR bands for this experiment is found in Table~\ref{table2}. The solid-state hydrogenation of an ice containing C$_{2}$H$_{2}$ leads to the formation of C$_{2}$H$_{4}$ and C$_{2}$H$_{6}$, which was also reported by \citet{kobayashi2017hydrogenation}. The reaction of CO and H, which has been extensively investigated by \citet{watanabe2002efficient} and \citet{fuchs2009hydrogenation}, yields H$_{2}$CO and CH$_{3}$OH. There is no clear spectral proof of propanal or 1-propanol.  

\begin{figure*}[btp!]
\includegraphics[width=15cm]{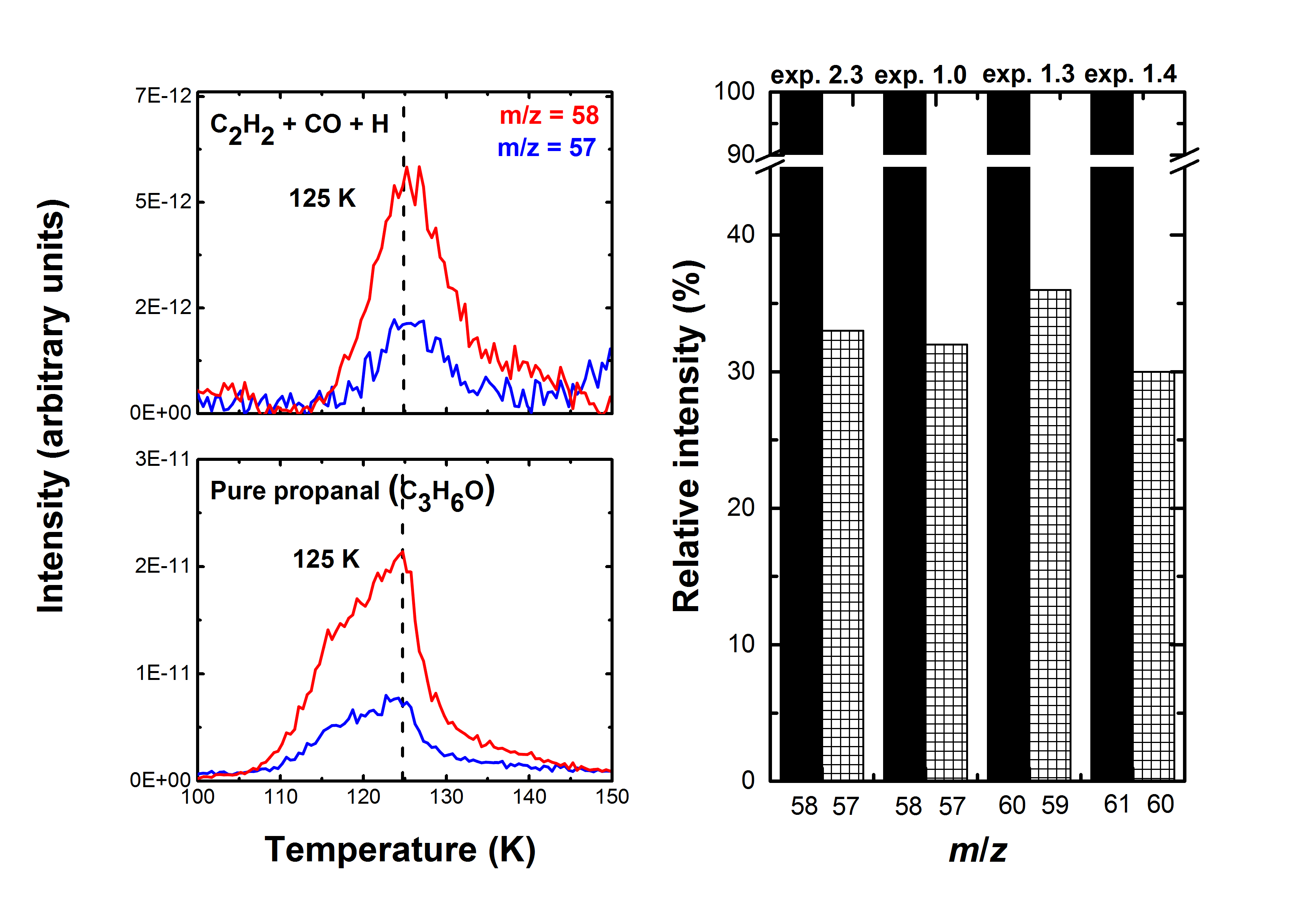}
\caption{(Left) TPD spectra of C$_{2}$H$_{2}$ + CO + H (top; exp. 1.0) and propanal (bottom; exp. 2.3) taken after deposition at 10 K. (Right) QMS fragmentation pattern of two \emph{m/z\/} values that are normalized to the QMS signal of the C$_{3}$H$_{6}$O$^+$ ion (or the corresponding isotopologue) found in the propanal (exp. 2.3), C$_{2}$H$_{2}$ + CO + H (exp. 1.0), C$_{2}$H$_{2}$ + C$^{18}$O + H (exp. 1.3), and C$_{2}$H$_{2}$ + $^{13}$C$^{18}$O + H (exp. 1.4) experiments.}
\label{fig4}
\end{figure*}

\begin{figure}[hbt!]
\includegraphics[width=\columnwidth]{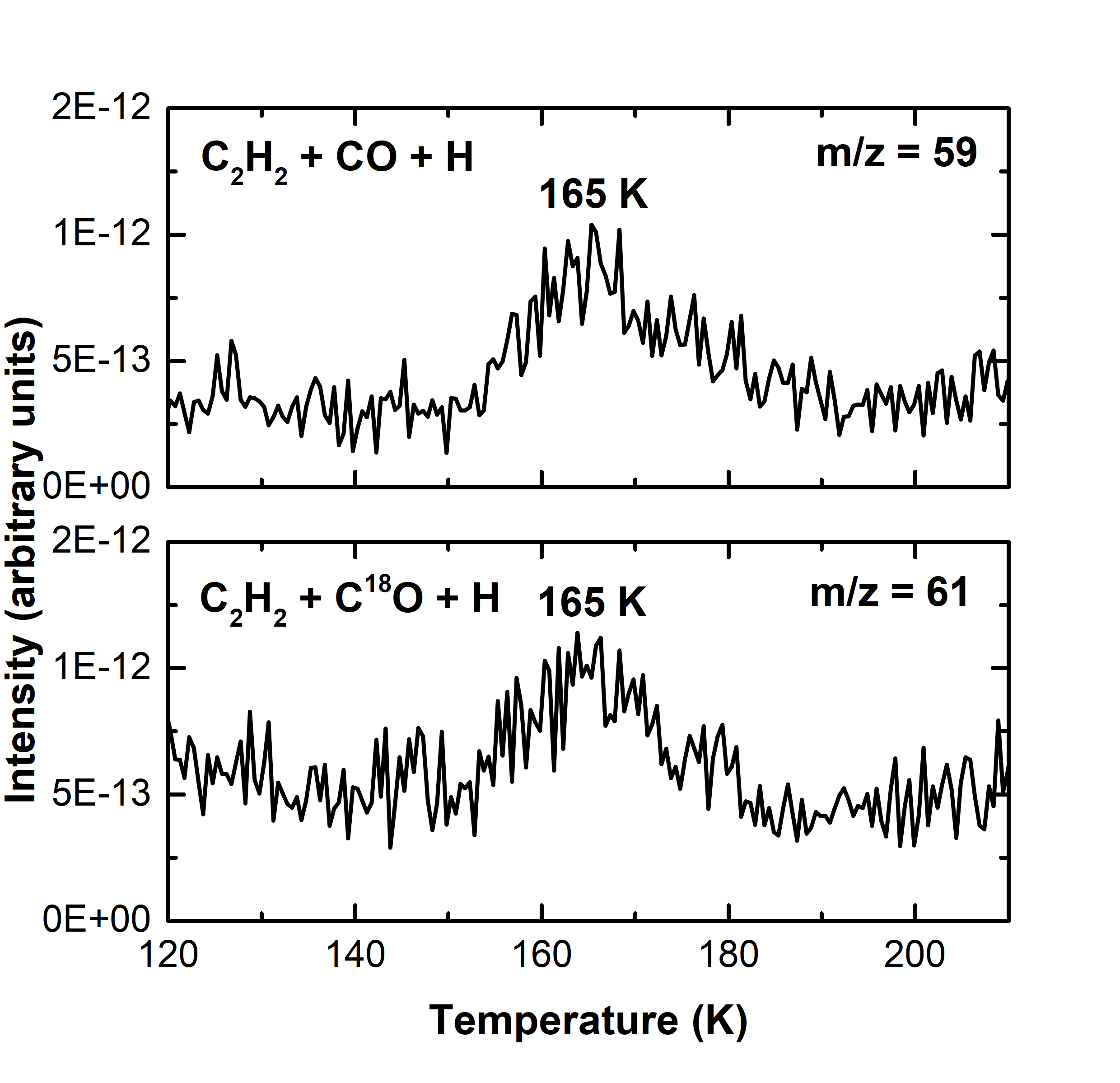}
\caption{TPD spectra that include \emph{m/z\/} values that may represent the desorption of 1-propanol. TPD of the reactions, C$_{2}$H$_{2}$ + CO + H (top; exp. 1.0) and C$_{2}$H$_{2}$ + C$^{18}$O + H (bottom; exp. 1.3), taken after deposition at 10 K.}
\label{fig5}
\end{figure}

\begin{figure*}[btp!]
\includegraphics[width=\textwidth]{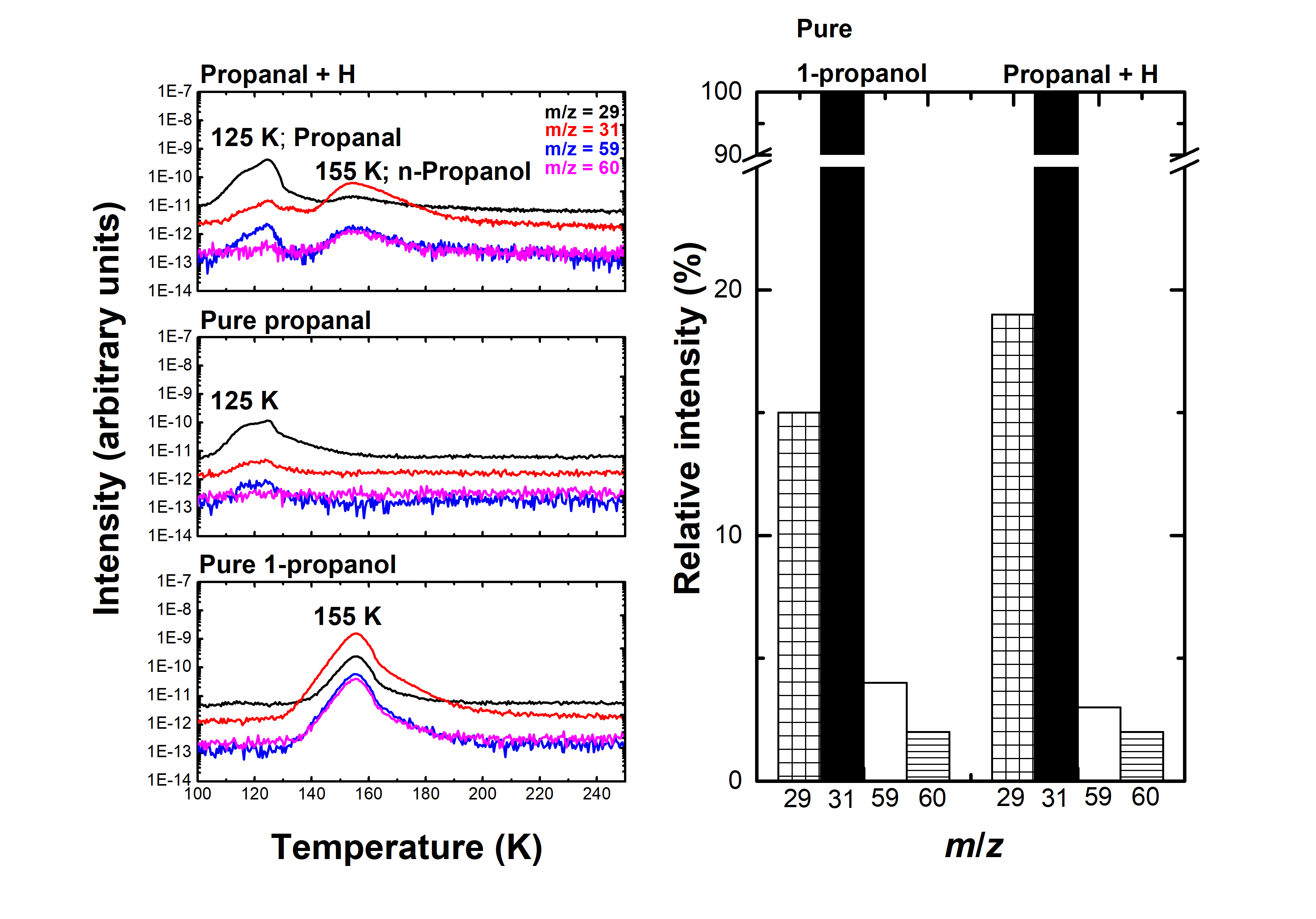}
\caption{(Left) TPD of propanal + H (top; exp. 2.2), propanal (middle; exp. 2.3), and 1-propanol (bottom; exp. 2.0) taken after deposition at 10 K. (Right) QMS fragmentation pattern of four \emph{m/z\/} values that are normalized to the QMS signal of \emph{m/z\/} = 31 found in the 1-propanol (exp. 2.0) and propanal + H (exp. 2.2) experiments for a temperature of 125 K.}
\label{fig2}
\end{figure*}  

\begin{figure}[hbt!]
\includegraphics[width=\columnwidth]{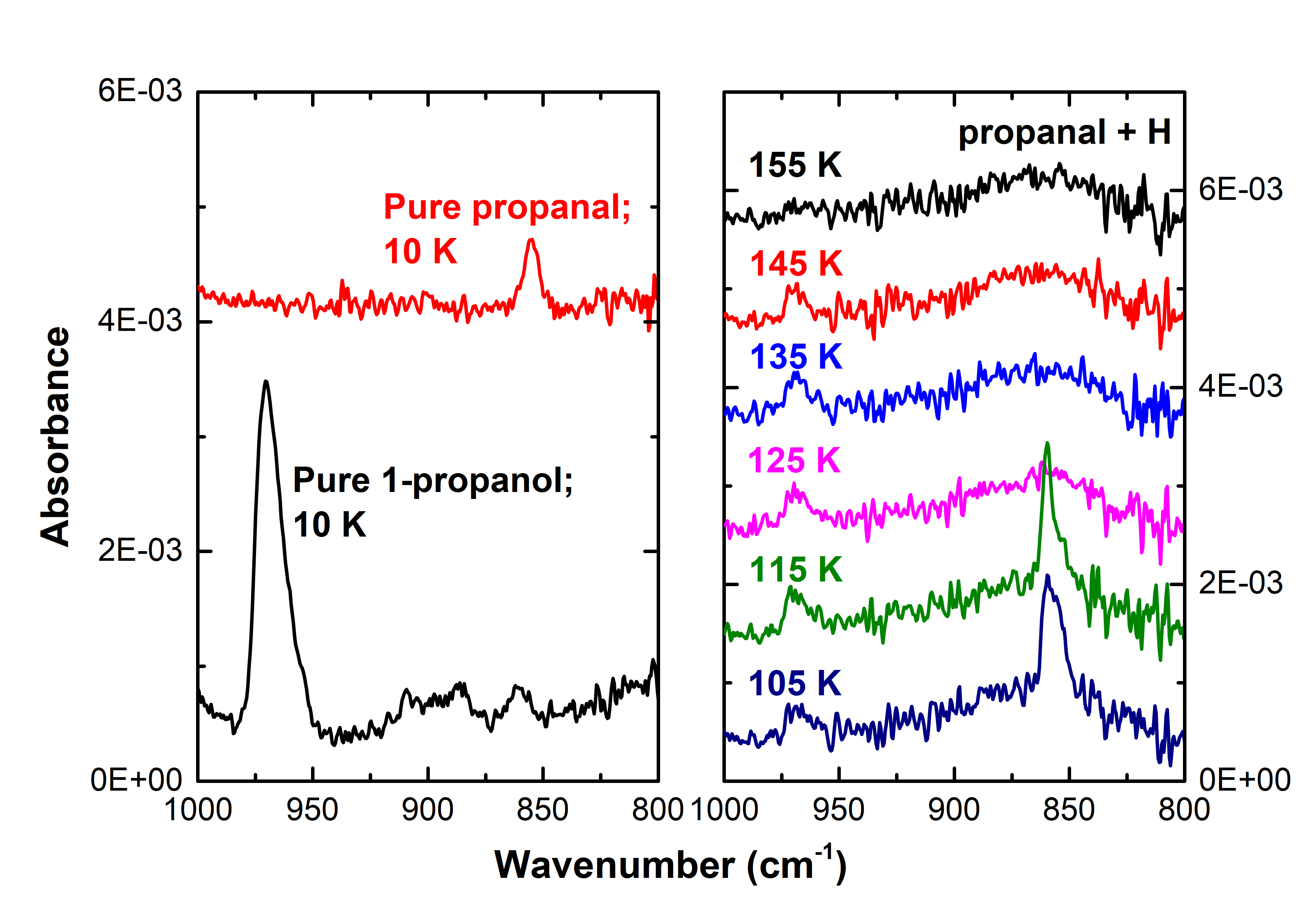}
\caption{(Left) Infrared features of pure propanal (exp. 2.3) and 1-propanol (exp. 2.0). (Right) RAIRS annealing series of propanal + H (exp. 2.1), taken after deposition at 10 K. We note that the features at 860 cm$^{-1}$ and 969 cm$^{-1}$ are signatures of propanal and newly formed 1-propanol (tentative), as the signatures disappear by 125 K (propanal peak desorption temperature) and 155 K (1-propanol peak desorption temperature), respectively. RAIR spectra are offset for clarity.}
\label{a2}
\end{figure}

Besides the resulting RAIR spectrum of C$_{2}$H$_{2}$ + CO + H in Figure~\ref{fig3}, also RAIR spectra of several control experiments are shown. The C$_{2}$H$_{2}$ + CO RAIR spectrum shows two features that belong to C$_{2}$H$_{2}$ and CO, but does not show the signatures of the other CH- and HCO-bearing species that are seen in the RAIR spectrum when H is present. As expected, this implies that H-atoms, and subsequently radicals, are required for the formation of C$_{2}$H$_{4}$, C$_{2}$H$_{6}$, H$_2$CO, and CH$_3$OH in the C$_{2}$H$_{2}$ + CO + H experiment. Some of the spectra of these reaction products are shown in Figure~\ref{fig3} to point out their IR features in the C$_{2}$H$_{2}$ + CO + H experiment.  
The RAIR spectra of pure propanal and 1-propanol in Figure~\ref{fig3} illustrate the obstacle of detecting these species as reaction products in the RAIRS data of the C$_{2}$H$_{2}$ + CO + H experiment. The strongest band of propanal overlaps with the feature of H$_{2}$CO ($\sim$1750 cm$^{-1}$), whereas the strongest bands of 1-propanol overlap with the features of C$_{2}$H$_{4}$ ($\sim$950 and $\sim$2950 cm$^{-1}$), C$_{2}$H$_{6}$ ($\sim$2950 cm$^{-1}$), and CH$_{3}$OH ($\sim$1050 cm$^{-1}$), as shown in Figure~\ref{fig3} by the dashed and dotted lines. With such closely overlapping features, even the incorporation of propanal and 1-propanol in a matrix containing relevant reactant species, which would affect the peak positions and profiles, would likely not lead to the explicit detection of propanal and 1-propanol IR signatures. Due to the lack of distinguishable IR peaks of propanal and 1-propanol in the C$_{2}$H$_{2}$ + CO + H spectrum, it is necessary to resort to an alternative detection method, such as TPD.

TPD spectra along with the QMS cracking pattern of synthesized and deposited propanal are compared in Figure~\ref{fig4}. In the TPD spectra obtained after the co-deposition of C$_{2}$H$_{2}$ + CO + H (top left), the \emph{m/z\/} signals of 58 and 57 peak at 125 K, which is what is observed in the TPD spectra of a pure propanal ice (bottom left). We note that there is a shoulder around 115 K in the pure propanal experiment that is not observed in the C$_{2}$H$_{2}$ + CO + H experiment. This is believed to be caused by the phase transition of propanal, which occurs during the desorption of propanal, as verified by the sharpening of the IR peaks in the RAIR spectra that are recorded at different temperatures (not shown here). Because propanal is mixed with other species in the C$_{2}$H$_{2}$ + CO + H experiment, it is much harder for these molecules to rearrange into the crystalline form, hence the lack of the phase transition shoulder in the top left figure. The fragmentation pattern involving the C$_{3}$H$_{6}$O$^{+}$ (\emph{m/z} = 58) and C$_{3}$H$_{5}$O$^{+}$ (\emph{m/z} = 57) ions that derive from propanal is shown (Fig.~\ref{fig4}, (right)) to complement the TPD findings. A fragmentation pattern of 33:100, 32:100, 36:100, and 30:100 is measured for the two ions from experiments 2.3, 1.0, 1.3, and 1.4, respectively. It is clear that the fragmentation pattern between the isotopically enhanced reactions is consistent and additionally their average value matches that of the pattern seen in the pure propanal experiment. The information from the discussed TPD experiments supports the hypothesis that propanal is formed in the C$_{2}$H$_{2}$ + CO + H experiment.

\begin{figure*}
\includegraphics[width=\textwidth]{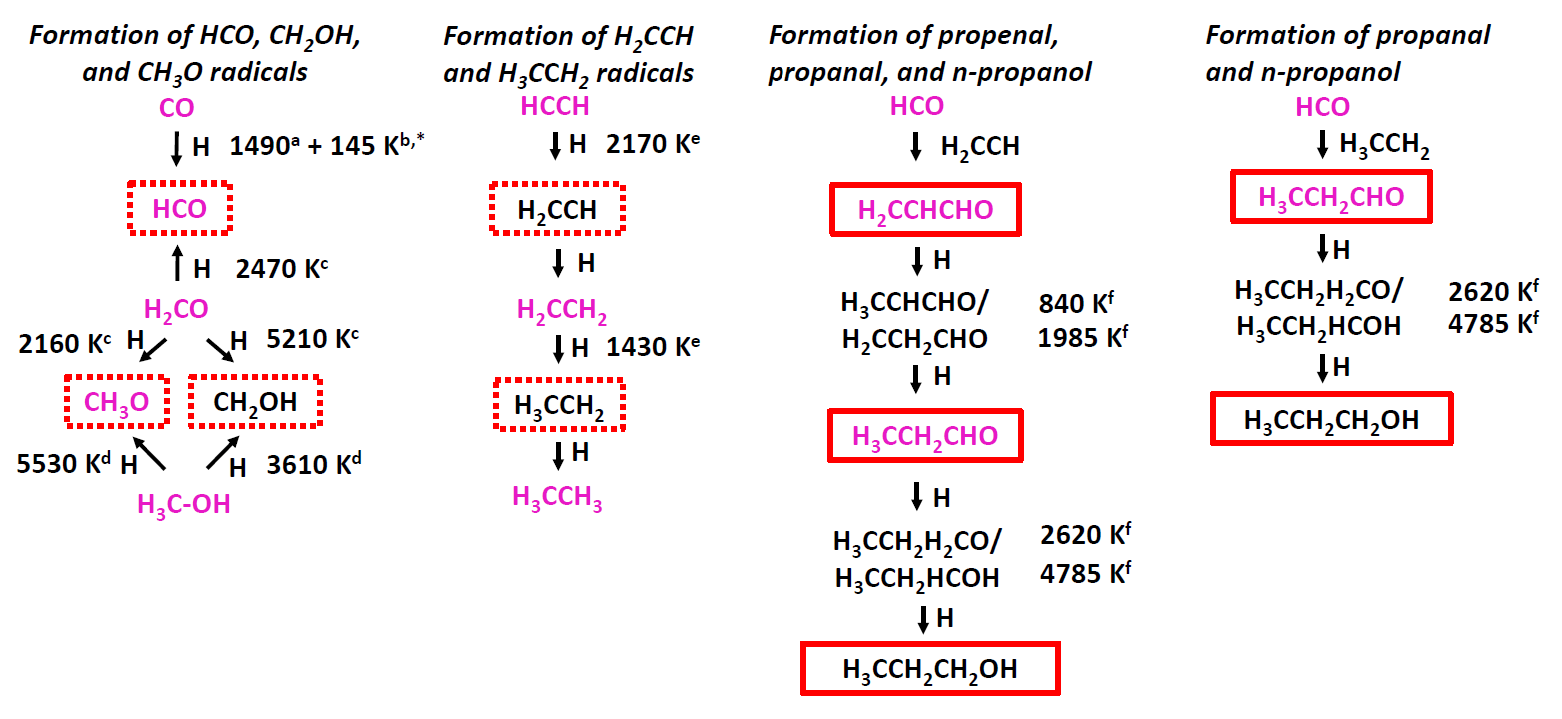}
\caption{Proposed mechanisms for experiment 1.0. We note that all radical--radical reactions shown here are barrierless. Relevant species within each mechanism are boxed; solid-line boxes indicate stable species and dotted-line boxes indicate radicals. Species labelled with purple font are those that have been detected in space. Activation energies are by a) \citet{andersson2011tunneling}, b) \citet{alvarez2018hydrogen}, c) \citet{song2017tunneling}, d) \citet{goumans2011deuterium}, e) \citet{kobayashi2017hydrogenation}, and f) \citet{zaverkin2018tunnelling}. An asterisk indicates the zero-point energy (ZPE) contribution.}
\label{figmech}
\end{figure*}

Due to the limited abundance of the formed propanal starting from C$_2$H$_2$ + CO + H and the desorption of side products that appear around the desorption of pure 1-propanol (e.g., glycolaldehyde), the detection of 1-propanol starting from a propanal-poor sample is just around the limit of our detection capabilities. Figure~\ref{fig5} shows TPD spectra of \emph{m/z\/} values that are tentatively identified as the C$_{3}$H$_{7}$O$^{+}$ and C$_{3}$H$_{7}$$^{18}$O$^{+}$ ions of 1-propanol. These \emph{m/z\/} values (59 and 61) are selected as they should not appear for glycolaldehyde desorption, which occurs already around 160 K. The peak desorptions at 165 K are shifted +10 K from the peak desorption temperature of pure 1-propanol (155 K), which can be explained by the desorption of 1-propanol from the bare substrate surface and/or sub-monolayer regime. In this case, molecules occupy spots with higher binding energies. Although the signal intensities between the two desorption peaks are similar and both \emph{m/z\/} values peak at the same temperature, more information (i.e., more \emph{m/z\/} channels) is needed to conclusively prove that 1-propanol formation can also be directly detected in the C$_2$H$_2$ + CO + H experiment. For this reason, we present results for the hydrogenation of propanal, which is shown in the following section. A similar two-step approach was used in a previous study to confirm the formation of glycerol from CO + H \citep{fedoseev2017formation}.      


\subsection{Formation of 1-propanol by solid-state hydrogenation of propanal}
\label{3.1} 

To confirm the formation of 1-propanol by solid-state hydrogenation of propanal ice, TPD spectra were collected and are presented in Fig.~\ref{fig2}. The TPD spectra of propanal + H, propanal, and 1-propanol for \emph{m/z\/} = 29, 31, 59, and 60 are displayed top-down in the left panel, as these \emph{m/z\/} values are representative of the ions produced when propanal and 1-propanol are fragmented by the QMS ionization source. For a pure propanal ice, the desorption peaks of \emph{m/z\/} = 29, 31, and 59 appear at 125 K, and are also found in the propanal + H experiment, as expected. In the propanal + H experiment, desorption peaks of \emph{m/z\/} = 29, 31, 59, and 60 appear also at 155 K, which are observed in the 1-propanol experiment. To confirm that the signals at 155 K in the propanal + H experiment are due to the desorption of 1-propanol ice, the fragmentation patterns of the \emph{m/z\/} values found in the propanal + H and pure 1-propanol experiments were compared (right panel). The relative intensities in the propanal + H experiment are 19:100, 3:100, and 2:100 for \emph{m/z\/} = 29:31, \emph{m/z\/} = 59:31, and \emph{m/z\/} = 60:31, respectively. These relative intensity values are almost identical to those found in the 1-propanol reference experiment, which are 15:100, 4:100, and 2:100 for these three \emph{m/z\/} values. This confirms that 1-propanol is derived from the hydrogenation of propanal at 10 K.   

To further complement the results from Fig.~\ref{fig2}, the formation of 1-propanol from the hydrogenation of propanal can be tentatively identified from the RAIRS annealing series (RAIR spectra recorded at different temperatures) presented in Fig.~\ref{a2}. The feature at 860 cm$^{-1}$ is assigned to the CH$_3$ rocking mode of propanal \citep{korouglu2015propionaldehyde} and the band at 969 cm$^{-1}$ overlaps nicely with the C-O stretching frequency of 1-propanol \citep{max20021}. As seen in the figure, the propanal band disappears at 125 K, which is in-line with the peak desorption temperature of 125 K for propanal, as demonstrated in Fig.~\ref{fig2}. The 969 cm$^{-1}$ feature disappears at 155 K, which is also the peak desorption temperature of 1-propanol. The results from Fig.~\ref{a2} provide additional evidence of 1-propanol formation from propanal + H, even though the figure only shows one potential band of 1-propanol. Other RAIR bands of 1-propanol cannot be positively identified or probed largely due to the low signal-to-noise ratio of the new bands in experiment 2.1. The data shown in Fig.~\ref{a2} support the results from the TPD experiments that are presented in Fig.~\ref{fig2}.          

\section{Discussion}
\label{4}

Figure~\ref{figmech} shows a list of possible pathways that hold the potential to form propanal and 1-propanol by the co-deposition of C$_{2}$H$_{2}$ + CO + H under our experimental conditions. These aim to mimic interstellar conditions as closely as possible, but one must bear in mind that mixed CO:C$_2$H$_2$ ices are likely not representative for interstellar ices. Here, we mainly aim at reproducing conditions that allow to study reaction pathways that will be at play in interstellar ices. The two left-most reaction chains in Fig.~\ref{figmech} show how the reacting radicals and stable molecules from the hydrogenation of CO (HCO, H$_2$CO, CH$_3$O, and CH$_2$OH) and C$_{2}$H$_{2}$ (H$_2$CCH, H$_2$CCH$_2$, and H$_3$CCH$_2$) are formed. We note that CO and C$_2$H$_2$ do not react with each other under our experimental conditions. From this set of radicals and molecules, the combination of which most likely leads to the formation of propanal and 1-propanol is discussed here first by process of elimination. The barrier value for H-abstraction from C$_2$H$_2$ is > 56,000 K \citep{zhou2008pathways}, which is very high for thermalized H-atoms to bypass at cryogenic temperatures used in our experiments. This H-abstraction is required for species -- such as propynal -- to be formed. Therefore, the pathways involving the formation of propynal are excluded from our reaction network. A direct consequence of this is that the C$\equiv$C bond must be converted to a single C-C bond by H-atom addition, as demonstrated in the works of \citet{hiraoka2000study} and \citet{kobayashi2017hydrogenation}.

Radical--molecule reactions, such as those between the HCO radical and C$_2$H$_2$ or C$_2$H$_4$ molecules, can also be excluded due to their high activation barriers. These activation energies are calculated following the method described by \citet{kobayashi2017hydrogenation} and \citet{zaverkin2018tunnelling}. Briefly, the electronic structure is described by density functional theory (DFT) with the MPWB1K functional \citep{zhao2004hybrid} and the def2-TZVP basis set \citep{weigend1998ri}. This combination has been shown to yield good results via benchmark studies. The activation energies are calculated including ZPE and with respect to the pre-reactive complex. Transition state geometries are listed in Table~\ref{table3} in Appendix~\ref{appendix}. These values are determined for the gas-phase, which will yield representative values as we expect the influence of the predominantly CO-rich environment to play a minor role in altering the reaction potential energy landscape. We find the activation energy for the reaction $\hbox{HCO + C$_2$H$_2$} \rightarrow \hbox{HCCHCHO}$ to be 4290 K and that for the reaction $\hbox{HCO + C$_2$H$_4$} \rightarrow \hbox{H$_2$CCH$_2$CHO}$ to be 3375 K. Such high barriers hint at a low overall efficiency, especially because, as indicated by \citet{alvarez2018hydrogen}, reactions where two heavy atoms are involved, for example formation of a carbon--carbon bond, are expected not to tunnel efficiently. Such barriers could be overcome if the HCO radical would have considerable leftover excess energy after formation.

With the exclusion of H-abstraction reactions involving stable hydrocarbon molecules and also radical--neutral reactions, the following reactions are left to consider: HCO + H$_2$CCH/H$_3$CCH$_2$, CH$_3$O + H$_2$CCH/H$_3$CCH$_2$, and CH$_2$OH + H$_2$CCH/H$_3$CCH$_2$. Of these, only HCO + H$_2$CCH/H$_3$CCH$_2$ leads to the formation of both -- propanal and 1-propanol. As shown in Fig.~\ref{figmech}, propenal can be formed by HCO + H$_2$CCH. Propenal was not detected in our experiments, and this is likely due to the low activation barrier of 842 K for propenal + H \citep{zaverkin2018tunnelling}, effectively converting propenal to further hydrogenation products. CH$_3$O and CH$_2$OH radicals may react with hydrocarbon radicals to form methoxyethene, methoxyethane, allyl alcohol, and 1-propanol. Yet, these radicals are further down the CO + H chain, and since H$_2$CO + H has a barrier of > 2000 K \citep{woon2002modeling,song2017tunneling}, reactions with CH$_3$O and CH$_2$OH radicals are less probable than with HCO under our experimental conditions. However, it should be noted that interstellar CH$_3$OH (ice and gas) is an abundant molecule that is primarily formed by the CO + H surface reaction, thus CH$_3$O and CH$_2$OH radicals must also be abundant in the ISM. Therefore other primary alcohols, aldehydes, and even ethers maybe formed with abundances that can be used to search for astrochemical links.

Comparison of the hydrogenation activation barriers of H$_2$CO and propanal shows that the values have a difference of < 500 K (with H$_2$CO + H having the smaller barrier), although the low-temperature rate constant is greater for the case of H$_2$CO. Since hydrogenation of H$_2$CO is the dominating pathway to CH$_3$OH formation in interstellar space, this means that also the hydrogenation of propanal resulting in the formation of interstellar 1-propanol maybe a notable pathway.

The work by \citet{jonusas2017reduction}, in which propanal hydrogenation was not found to result in 1-propanol formation, seems to be in contradiction with our findings. A direct comparison is hard, since the hydrogen and propanal fluxes and fluences, and particularly the deposition methods, are different between the two studies. \citet{jonusas2017reduction} deposited propanal first, then bombarded the ice with hydrogen atoms. This is known as the pre-deposition method, which results in less product formation in comparison to the co-deposition method usually because of the limited penetration depth of hydrogen atoms in the ice, as discussed by \citet{fuchs2009hydrogenation} in the case of CO + H. The theoretical work by \citet{zaverkin2018tunnelling} suggested that the non-detection of 1-propanol by \citet{jonusas2017reduction} could be due to the continuous H-abstraction and subsequent H-addition from and onto the carbonyl-C, respectively, since H-abstraction from the carbonyl-C of propanal was found to be five orders of magnitude faster than H-addition to O at 60 K (we note that the experiments presented here occur at 10 K). Another scenario could exist: after H-abstraction from the carbonyl-C, the resulting radical could be more prone to hydrogenation on the O, which would favour 1-propanol formation. However, there are no rate constants or branching ratios available for that process.

Finally, we address the dominant reaction mechanism. Reactions that take place on surfaces such as that studied here usually have three mechanisms: Langmuir-Hinshelwood (L-H), Eley-Rideal (E-R), and hot-atom (H-A) \citep{he2017mechanism}. In the presented experiments, the ice temperature is at 10 K during the deposition. This allows the residence time of H-atoms to be long enough for the atoms to rapidly scan the surface and have multiple chances of reaction with other ice reactants. Further, the rate of reaction via the L-H mechanism dominates over E-R and H-A mechanisms especially when the reaction possesses a significant activation barrier. As demonstrated by \citet{watanabe2002efficient}, \citet{watanabe2003dependence}, \citet{cuppen2007simulation}, \citet{fuchs2009hydrogenation}, \citet{chuang2015h}, and \citet{qasim2018formation}, the abundance of products that are formed from hydrogenation decreases substantially as the deposition temperature increases to temperatures that are below the initial desorption temperature of the reactant molecule(s). This is due to the rapid drop of the H-atom residence time on the surface. If the E-R or H-A mechanism were responsible for the formation of products, then no drastic drop in the amount of the formed products would be observed. This evidence in favour of the L-H mechanism also allows us to claim that the H-atoms involved in the reactions are in thermal equilibrium with the 10 K surface. 

\begin{figure*}
\includegraphics[width=\textwidth]{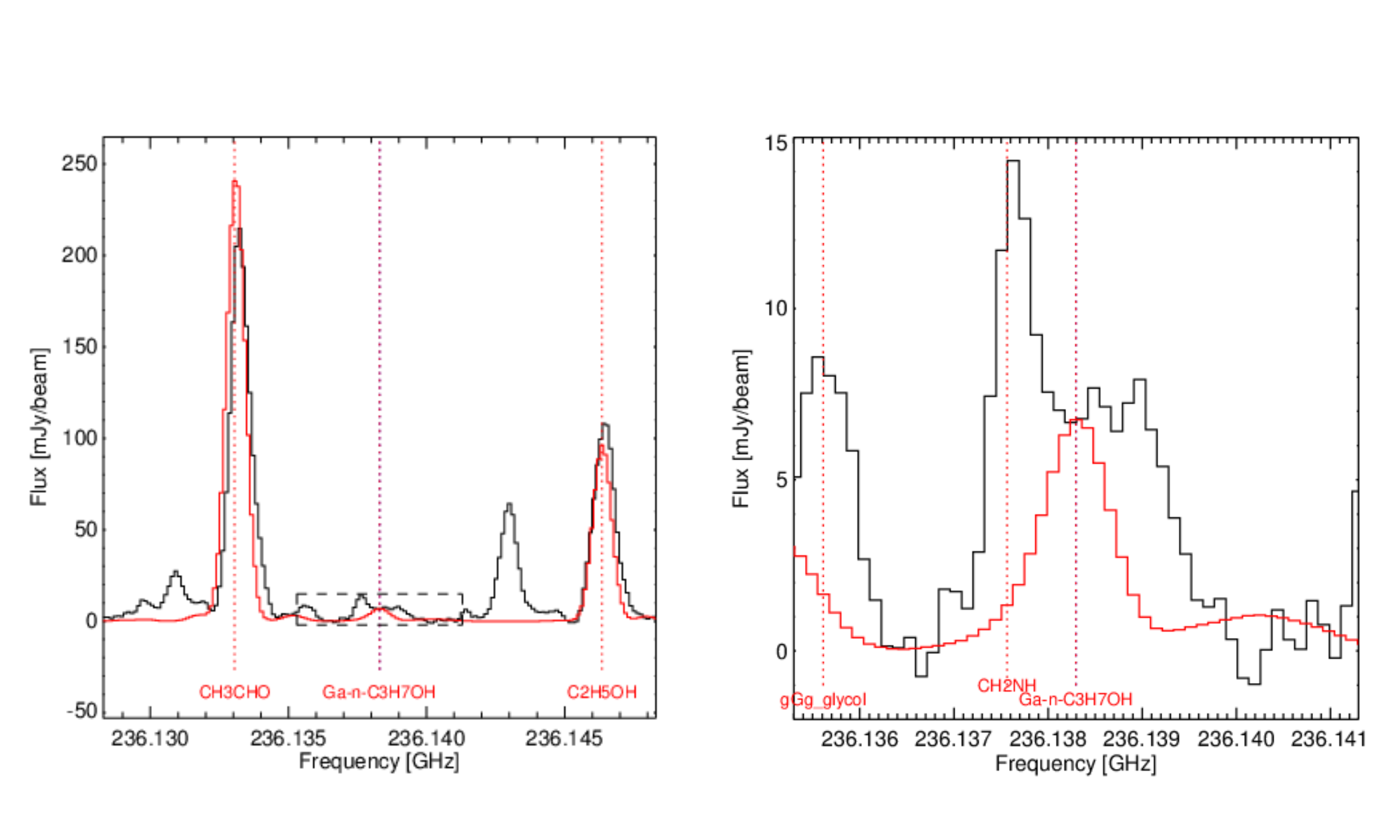}
\caption{Extended (left) and zoomed-in (right) spectra around the 1-propanol transition. Observed spectrum (black) around the targeted 1-propanol transition at 236.138 GHz (purple dotted-line and black dashed-line box) towards the ``full-beam" offset position located 0.5\arcsec away from the continuum peak of IRAS 16293-2422B. Synthetic spectrum of the LTE model is shown in red. The predicted 1-propanol transition shown here is for $N$(1-propanol) = $1.2 \times 10^{15}$ cm$^{-2}$ and $T_{\rm ex} = 300$ K (see text for more details). Red dotted-lines refer to the position of transitions of identified species detected above 5$\sigma$, with the associated species labelled below the spectrum.
}
\label{fig6}
\end{figure*}

\section{Astrophysical implications}
\label{astro}

The experimental conditions and chemical species studied aim to mimic reaction pathways that can take place on icy dust grains in a cold and dense prestellar core or the outer regions of protostellar envelopes (i.e., 10 K ices formed primarily by radical-induced reactions). Specifically, we have investigated how species formed along the well-studied CO hydrogenation chain can interact with radicals formed upon hydrogenation of other species expected to be present in an interstellar ice environment. Newly formed ice constituents can then be observed in the gas-phase after warm-up in the hot core region following thermal desorption. Following the outcome of our experiments, the detection of propanal in hot cores may be explained following the reaction scheme discussed in Fig.~\ref{figmech} and the formation of 1-propanol is a logical consequence, providing solid motivation for future surveys for this species. C$_2$H$_2$ was used in the experiments as a source for hydrocarbon radicals, which are species that can also be formed in different ways in the ISM. Strong lines of gaseous C$_2$H$_2$ have been detected in warm gas in protostellar envelopes \citep{lacy1989discovery,lahuis2000iso,rangwala2018high} and in protoplanetary disks \citep{gibb2007warm,carr2008organic,salyk2011infrared}, with typical abundances of 10$^{-7}$ - 10$^{-6}$ with respect to H$_2$, or 10$^{-3}$ - 10$^{-2}$ with respect to gaseous H$_2$O or CO. However, there has not yet been a detection of interstellar solid C$_2$H$_2$. The limits on C$_2$H$_2$ ice are < 1.4\% with respect to H$_2$O ice \citep{boudin1998constraints}, which is similar to or lower than the abundance of CH$_4$ ice (typical abundance of $\sim$5\%) \citep{gibb2004interstellar,oberg2008c2d,oberg2011ices,boogert2015observations}. Other models of gas-grain chemistry predict lower C$_2$H$_2$ abundances; a factor of 50 - 100 lower than that of CH$_4$ \citep{garrod2013three}. In cometary ices, C$_2$H$_2$ is detected, at a level of 0.1 - 0.5\% with respect to H$_2$O ice \citep{mumma2011chemical}. A logical explanation for such low abundances is that the bulk of the solid C$_2$H$_2$ is transformed to other species, through reactions such as those studied here.   

As stated in Sect.~\ref{introduction}, 1-propanol has not yet been identified in the ISM, but several surveys have attempted its detection. Here we put the laboratory and theoretical findings presented in the previous sections into an astrochemical context, using deep interferometric observations by ALMA with the aim to constrain the abundance of 1-propanol around the hot core of the low-mass protostar IRAS 16293-2422B. We use the 12m array ALMA data from the work by \citet{taquet2018linking} under Cycle 4 (program 2016.1.01150.S) in Band 6 at 233 - 236 GHz. These observations have a circular Gaussian beam fixed to 0.5'' and with a 1$\sigma$ rms sensitivity of 1.2 - 1.4 mJy beam$^{-1}$ per 0.156 km s$^{-1}$ channel. This provides one of the deepest ALMA datasets towards a low-mass protostar obtained so far. Spectra of the four spectral windows obtained towards a position located at 1 beam size offset in the southwest direction with respect to the source B dust continuum position are analysed, which gives the best compromise between intensity and opacity of the continuum and the molecular emission. The observed and predicted spectra of the four spectral windows towards the full-beam offset position are shown in the Appendix of \citet{taquet2018linking}. As explained there, more than 250 spectroscopic entries mostly using the CDMS and JPL catalogues have been taken into account to identify all detected transitions. However, as discussed by \citet{taquet2018linking}, $\sim$70\% of the $\sim$670 transitions remain unidentified at a 5$\sigma$ level. The full spectrum of 1-propanol over the entire frequency range is simulated (Fig.~\ref{aaa} in Appendix~\ref{appc}) and compared with observations. The spectroscopic data of the 1-propanol molecule are provided by \citet{kisiel2010determination}. About 60 ``bright" transitions (i.e. $E_{\rm up} < 500$ K, $A_{\rm i,j} > 10^{-5}$ s$^{-1}$) from 1-propanol are located in the frequency range covered by the four spectral windows. The transition that gives the deepest constraint on the column density of 1-propanol is that at 236.138 GHz ($E_{\rm up} = 160$ K, $A_{\rm i,j} = 6.6 \times 10^{-5}$ s$^{-1}$) as seen in Fig.~\ref{aaa}. 

We derive the upper limit of the 1-propanol column density assuming conditions at the Local Thermal Equilibrium (LTE) and assuming optically thin emission and excitation temperatures of 300 and 125 K, following previous ALMA observations of other COMs towards this source \citep{jorgensen2018alma}. Both panels in Fig.~\ref{fig6} show the spectrum around the targeted transition obtained after a baseline correction through a fit over the line-free regions around 236.138 GHz. We note that only the spectrum at $T_{\rm ex} = 300$ K is shown, since the spectrum for $T_{\rm ex} = 125$ K at around 236.138 GHz is the same. The 1-propanol transition is blended by two lines at 236.1376 and at 236.1390 GHz, which is clearly visible from the zoom-in shown in the right panel. The former transition (on the left) could be partially attributed to CH$_2$NH, recently detected toward IRAS 16293-2422B by \citet{ligterink2018alma} using ALMA. The peak on the right is of unknown nature and may be due to a rotational transition starting from a vibrationally excited species. With an offset of 0.15 MHz with respect to the synthetic transition (red), it is unlikely that this peak is actually due to 1-propanol. Only a modification of the source velocity from 2.7 km/s -- the source velocity of IRAS16293-B usually derived -- to 2.5 km/s would result in a match. In that case, the next strongest transitions should be searched for. We verified that other ``bright" 1-propanol lines are not detected in our observed spectrum for the two different upper limits and associated excitation temperatures. For the moment, we conclude that the transition to the right is not due to 1-propanol.

In order to derive a conservative limit for the 1-propanol column density, we neglect the spectral contribution of the two peaks shown in Fig.~\ref{fig6} near the wavelength of the predicted 1-propanol transition and instead derive the column density using the synthetic transition. At 300 and 125 K, 1-propanol column densities of $1.2 \times 10^{15}$ cm$^{-2}$ and $7.6 \times 10^{14}$ cm$^{-2}$ are derived, respectively, which are the highest column densities that still result in a non-detection of 1-propanol. Comparing this value to the propanal column density of $2.2 \times 10^{15}$ cm$^{-2}$ found by \citet{lykke2017alma} for 125 K with similar observational properties, this results in a 1-propanol:propanal upper limit of $< 0.55$ ($T_{\rm ex} = 300$ K) and $< 0.35$ ($T_{\rm ex} = 125$ K). This is consistent with the experiments in this work and also with the theoretical calculations by \citet{zaverkin2018tunnelling}, which show that the hydrogenation of propanal to 1-propanol involves a barrier. From the perspective that only the activation barrier is considered, there should be less 1-propanol in space in comparison to propanal if 1-propanol originates from propanal.   

The C$_2$H$_2$ + CO + H experiment shows the importance of introducing different molecules to the CO + H channel. The CO hydrogenation chain is generally taken as the way to explain the observed CH$_3$OH abundances in space under dark cloud conditions. In recent work, an extension of this network towards larger sugars and sugar alcohols was proven. Here we demonstrate that this reaction chain also holds potential for the formation of other species, including radicals formed by other means. By adding C$_2$H$_2$, reaction pathways are realised in which 1-propanol can be formed. This is significant, as the molecule has astrobiological relevance and may already be formed during the dark cloud stage, for example when particularly `non-energetic' processes are at play. It is clear from the detections and proposed list of mechanisms in this work that the extension of the CO + H channel is promising to explain the formation of potentially important interstellar species that have solid-state formation pathways that are not yet well understood.

From the studied reactions, it can be generalized that a whole set of various aldehydes and primary alcohols can be formed starting from CO and polyynes, where polyynes are composed of alkynes such as C$_2$H$_2$. Such molecules can directly participate in the formation of micelles, or serve as the analogues of fatty acids in the formation of glycerol esters (analogues of glycolipids). The latter is particularly intriguing since previous results indicate that glycerol is formed by hydrogenation of CO during the heavy CO freeze-out stage \citep{fedoseev2017formation}.  

\section{Conclusions}

\label{conclusions}

This study focuses on the possible formation of the COMs, propanal and 1-propanol, that may take place when radicals formed in the hydrogenation of C$_2$H$_2$ and CO ice interact. For a temperature of 10 K and upon H-atom addition during a C$_2$H$_2$ and CO co-deposition experiment, our findings can be summarised as follows.

\begin{itemize}
  \item We find the formation of propanal and possibly 1-propanol ice. 
  
  \item We show that the hydrogenation of propanal ice leads to 1-propanol formation. Further theoretical investigations on the scenario that favours 1-propanol formation are desired. 
  
  \item We conclude that the most likely formation scheme of these two COMs is through the radical--radical reactions of HCO + H$_2$CCH and HCO + H$_3$CCH$_2$.
  
  \item We derive 1-propanol upper limits of $1.2 \times 10^{15}$ cm$^{-2}$ ($T_{\rm ex} = 300$ K) and $7.6 \times 10^{14}$ cm$^{-2}$ ($T_{\rm ex} = 125$ K) from ALMA observations towards the IRAS 16293-2422B low-mass protostar. These values are compared to the propanal column density of $2.2 \times 10^{15}$ cm$^{-2}$ from \citet{lykke2017alma}. The 1-propanol:propanal abundance ratio of $< 0.35 - 0.55$ is complemented by activation barriers of $\hbox{propanal + 2H} \rightarrow $\hbox{1-propanol} found in the presented experiments and in theoretical works.
\end{itemize}

\begin{acknowledgements}
This research would not have been possible without the financial support from the Dutch Astrochemistry Network II (DANII). Further support includes a VICI grant of NWO (the Netherlands Organization for Scientific Research) and A-ERC grant 291141 CHEMPLAN. Funding by NOVA (the Netherlands Research School for Astronomy) and the Royal Netherlands Academy of Arts and Sciences (KNAW) through a professor prize is acknowledged. D.Q. thanks Johannes K{\"a}stner for insightful discussions. G.F. and V.T. recognise the financial support from the European Union's Horizon 2020 research and innovation programme under the Marie Sklodowska-Curie grant agreement n. 664931. T.L. is supported by NWO via a VENI fellowship (722.017.00). S.I. recognises the Royal Society for financial support and the Holland Research School for Molecular Chemistry (HRSMC) for a travel grant. This paper makes use of the following ALMA data: ADS/JAO.ALMA\#2016.1.01150.S. ALMA is a partnership of ESO (representing its member states), NSF (USA) and NINS (Japan), together with NRC (Canada), MOST and ASIAA (Taiwan), and KASI (Republic of Korea), in cooperation with the Republic of Chile. The Joint ALMA Observatory is operated by ESO, AUI/NRAO and NAOJ.
\end{acknowledgements}
\bibliography{propanol}

\clearpage

\appendix

\makeatletter
\def\fps@figure{htbp!}
\def\fps@table{htbp!}
\makeatother

\onecolumn
\begin{multicols}{2}
\section{Additional RAIR spectra}

\includegraphics[width=\columnwidth]{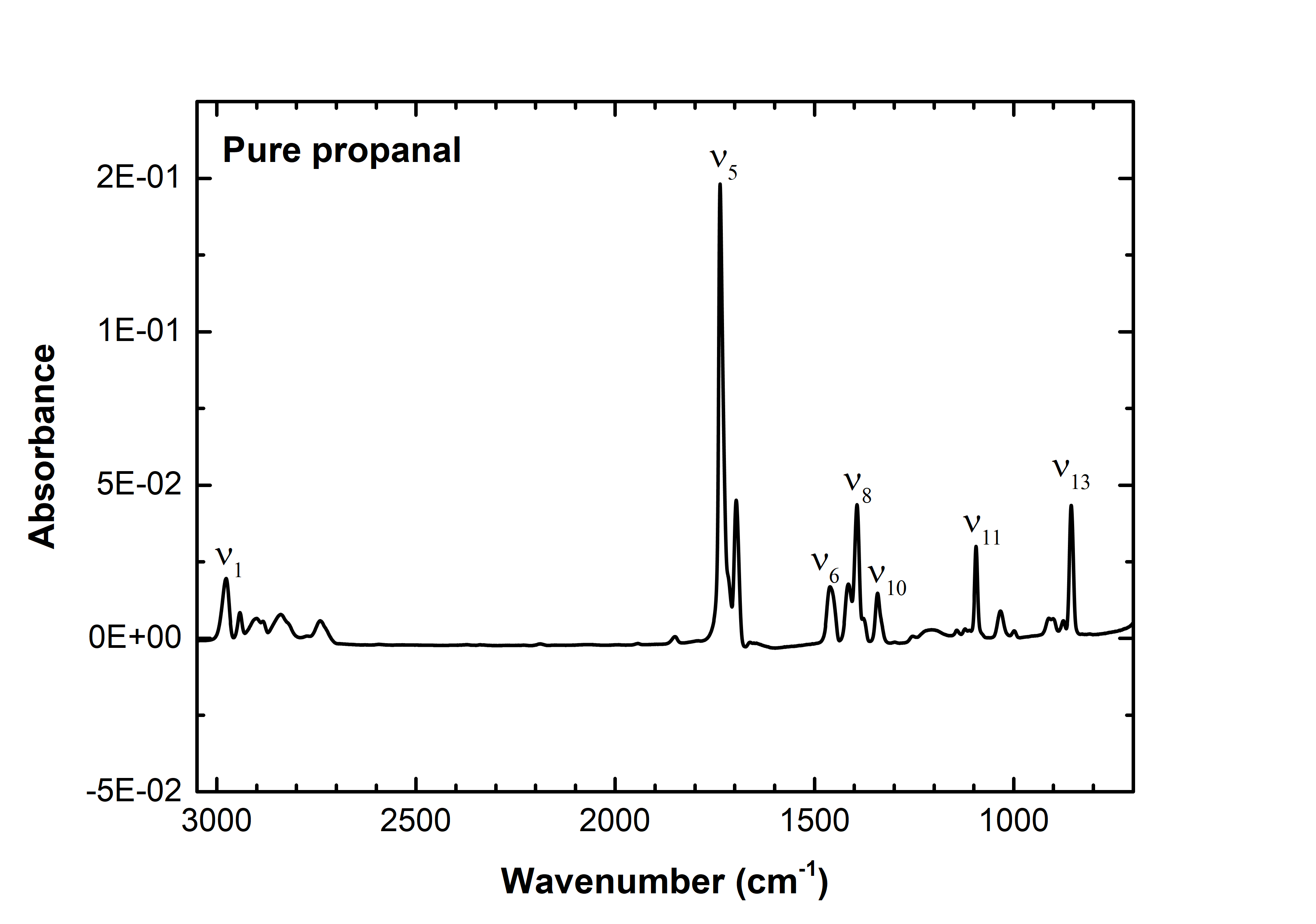}
\captionof{figure}{RAIR spectrum of propanal (exp. 2.4) taken at 10 K. Vibrational mode assignments are acquired from the work by \citet{korouglu2015propionaldehyde}.}
\label{a1}
\vspace*{1cm}

\columnbreak
\section{xyz coordinates of the transition state structures for HCO + C$_2$H$_2$ and HCO + C$_2$H$_4$}
\label{appendix}

    \small  
	\captionof{table}{Transition state (TS) geometries for HCO + C$_2$H$_2$ and HCO + C$_2$H$_4$ in the gas-phase.}
	\label{table3}
	\begin{tabular}{c c c c c c c c c c c c c c c c} 
		\hline
		& & TS &\\ 
		\hline
        & & R1: $\hbox{HCO + C$_2$H$_2$} \rightarrow \hbox{HCCHCHO}$ &\\
		C & 2.457338 & 0.140746 & 0.005500\\ 
		C & 2.631181 & -0.091562 & 1.180635\\ 
        H & 2.643330 & -0.284788 & 2.221672\\ 
        H & 2.726948 & 0.284977 & -1.012337\\ 
		H & -0.004679 & 0.618071 & 0.718368\\ 
        C & 0.378597 & 0.141577 & -0.202373\\ 
        O & -0.111318 & -0.735052 & -0.794637\\ 
        \hline
        & & R2: $\hbox{HCO + C$_2$H$_4$} \rightarrow \hbox{H$_2$CCH$_2$CHO}$ &\\
        C & 2.503771 & 0.131055 & -0.104025\\ 
        C & 2.587871 & -0.106220 & 1.216301\\ 
        H & 2.638529 & 0.697532 & 1.930552\\ 
        H & 2.561540 & -1.108352 & 1.607866\\ 
        H & 2.634615 & 1.125558 & -0.495603\\ 
        H & 2.576039 & -0.671720 & -0.817177\\ 
        H & 0.014387 & 0.814221 & 0.535546\\ 
        C & 0.362966 & 0.096368 & -0.230237\\ 
        O & -0.158186 & -0.904331 & -0.528368\\ 

		\hline
	\end{tabular}
\end{multicols}    

\vspace*{-2\baselineskip}
    
\section{1-propanol spectra at $T_{\rm ex}$ = 125 and 300 K}
\label{appc}
    
\begin{figure}[hbt!]\centering
\includegraphics[width=.92\textwidth]{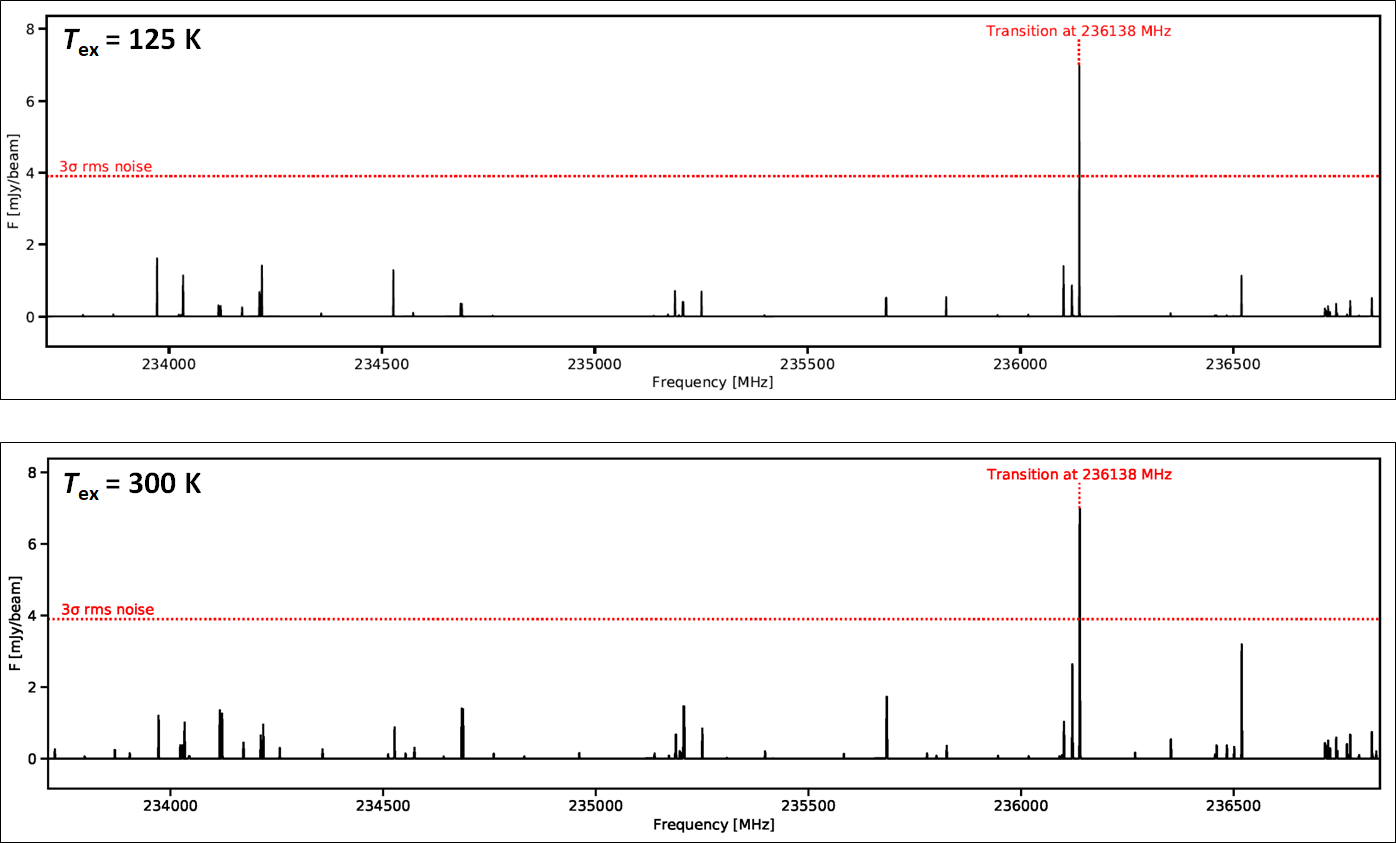}
\captionof{figure}{Synthetic spectra of the 1-propanol emission for excitation temperatures $T_{\rm ex} = 125$ K (top) and 300 K (bottom) with associated 1-propanol column densities of $7.6 \times 10^{14}$ cm$^{-2}$ and $1.2 \times 10^{15}$ cm$^{-2}$, respectively. These are the highest column densities that result in non-detection of the 1-propanol transition at 236.138 GHz (see text for more details).}
\label{aaa}
\end{figure}

\end{document}